A.M. Makhankin, V.V. Myalkovsky, V.D. Peshekhonov, S.E. Vasilyev

**Direct Timing Method for Longitudinal Coordinate Determination in Straws**

1. Introduction

Straws – thin-walled drift tubes (TDT) – are used for high-accuracy determination of radial coordinate of the nearest to the anode point of the particle traversal path through it. In planar detectors additional detecting planes with TDT inclined by a certain angle are usually used for determination of the other coordinate. However, in some cases it's preferable to have a possibility of a two-coordinate read-out for each detecting plane. For example, the barrel detector being cylindrical in shape can't be compact in regard to thickness with straws set in its detecting layers to different angles, which is often required.

To determine the longitudinal coordinate, a charge-division technique (CDT) with signal registration on both straw ends applies. However, the resolution for minimum ionizing particles (MIP) with this method is not high enough.

Readout of the longitudinal coordinate can also be made by a Constant-Fraction Timing (CFT).

Previously a principal possibility has been shown to determine the longitudinal coordinate using the method of measuring the difference in pulse arrival time at the two ends of drift tubes, so-called Direct Timing Method (DTM) [1]. A set of measurements has been carried out, which revealed advantages of this method [2], but the coefficient 2.5 has not been considered because of technical error. Besides, in research an artificial method was applied – shifting one pulse by cable delay to align it with the second one in order to determine the difference between times of their arrivals to amplifiers, which is possible in bench tests only.

The fulfilled works, which excluded technical error and allowed obtaining measurement results excluding the artificial method of difference determination between times of pulses' arrival at the two ends of the anode tubes, have been presented in [3]. This paper presents in brief the main results as well as describes several different ways of obtaining the results.

2. Principle of DTM

The longitudinal readout was studied for straws with a diameter of 9.53 mm and a length of 2 m, flashed with $Ar/CO_2$ (80/20) gas mixture under its pressure in the tube of 1 or 3 bar. A 30 µm diameter wire with resistance of 70 Ω·m served as an anode and an impedance of the straw was ~ 360 Ω. The straw was irradiated by gamma-ray quanta of a Fe-55 source with energy of 5.9 keV or by electrons from a $^{106}$Ru source with energy of 3.55 MeV. The pulses from the two ends of the anode were registered by two identical MSD-2 microcircuit-based amplifiers with the following inherent parameters: a gain of 35 mV/µA, a rise time of ~ 4 ns and an input impedance of 120 Ω. These amplifiers were developed and used for radial coordinate determination by measuring drift time of ionization electrons [4].
The pulses, passing along the anode with velocity v=3.49 ns/m, arrived at the amplifiers and then are led into two channels of a DRS4 amplitude-digital converter, which digitized them with a frequency of 2 or 5 GHz [5]. Then the data on pulse amplitude and shape were sent to the computer.

If the point of the avalanche origin is displaced along the anode by δL value from the center of the straw with L length, the two signals arriving at its amplifiers pass the distances (L/2 ± δL). Consequently, the time difference of the pulses δt determines the longitudinal coordinate relative to the straw center with the equation δt = ± 2δL / v, where v is a wave propagation velocity along the anode wire and the sign indicates in what direction it is moved from the center.

### 3. Results of the measurements with the DTM

#### 3.1 Measurements with the use of a cable delay

Using CDT in measurement of the longitudinal coordinate in the drift tubes gives a spatial resolution value for Fe-55 gamma-ray quantum registration, which is considerably higher than for minimum ionizing particle registration [6], while DTM provides closer values for both cases.

The time difference between pulse arrivals δt was measured as follows. A pair of correlated pulses from irradiated straw was chosen and the first pulse was shifted by calibrated cable delay in order to align it with the second one with accuracy of at least ± 300 ps. For both pulses a linear approximation of their leading edges was used in time range between top $T_t$ and bottom $T_b$ levels from pulse height to transverse with the DRS4 time axis in units of its bin. Value of δt was defined as a sum of cable delay value, which provided time pulse coincidence with difference in results of the linear approximation of pulse leading edges. Obtained values δt at a fixed source position towards the anode center were histogrammed by δL value and used for determination of average value and σ of longitudinal resolution.

The best magnitudes of the method accuracy were obtained with proper termination of amplifiers with TDT, for which 240 Ω series resistors were set at the amplifier inputs. During the measurements the values $T_b$ and $T_t$ levels were optimized for each position of the source separately. It should be noted that 1 cm of distance along the anode corresponded to time delay difference δt = 69.8 ps. The σ value of spatial resolution obtained at registration of gamma quanta from the $^{55}$Fe source at certain level values is shown in Fig.1-a. This figure demonstrates the longitudinal resolution dependence on levels of leading edge approximation. However, it can be expected that with the optimum match of levels and the avalanching point shifting from the tube center to its end the average values of longitudinal resolution range from ~70 to ~ 130 ps. Close values (Fig. 1-b) for $T_b$ and $T_t$ levels with values of 0.2 and 0.7, respectively, were defined by registration of high-energy electrons from $^{106}$Ru source. Fig. 1 is taken from [2]. The longitudinal resolution value ratio, determined by the DRS4 bin value of 500 ps, is shown on the Y-axis.

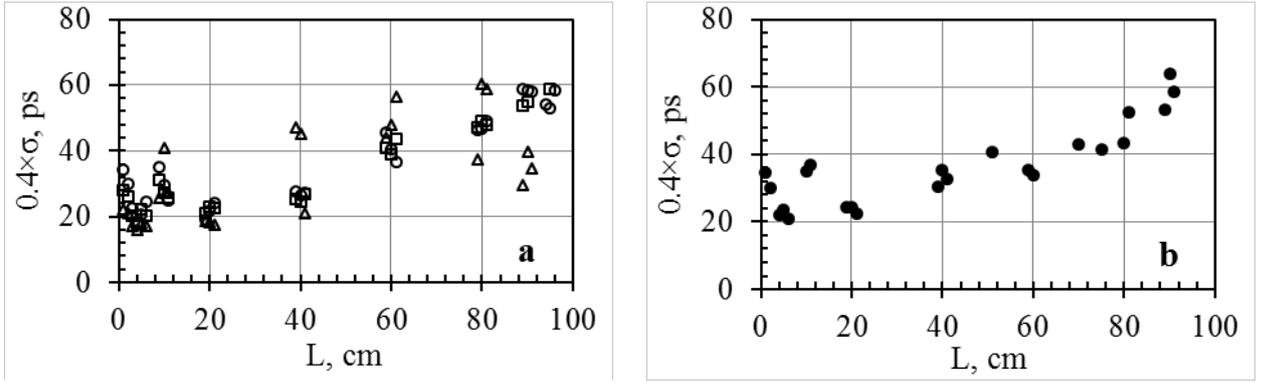

Fig. 1. The spatial resolution along one half of the straw. Gas mixture at a pressure of 1 bar and gas gain ~ $8 \times 10^4$.

a) - Registration of gamma rays with energies of 5.9 keV. Levels of $T_b$ and $T_t$ are 0.2 and 0.7 (open circle), 0.05 and 0.8 (open triangle), 0.1 and 0.6 (open square), respectively.

b) - Registration of electrons with the energy of 3.55 MeV. The levels of $T_b$ and $T_t$ are 0.2 and 0.7, respectively.

### 3.2. Measurement without cable delay

The time difference between correlated pulse arrivals δt was measured without cable delay according to the algorithm schematically shown in Fig.2. Correlated pulses were chosen by the setting of a $T_{min}$ threshold for the incoming anode pulses, and in the case of $^{106}$Ru source also by the trigger pulses sent to the DRS4 channel from the scintillation counter.

After the maximum values of these pulse pairs were found in a programmed time gate $t_s$ and the delayed pulse was normalized to the maximum amplitude of the first pulse the linear approximation of the pulse leading edges between top $T_t$ and bottom $T_b$ levels programmed in units of their maximum pulse values was made.

Then the pulse arrival times ($t_1$ and $t_2$) were defined by the cross-point of the approximation line with the time axis in units of clock generator (bins) of 200 and 500 ps at a frequency of 5 or 2 GHz, respectively. After that, values of δt time delay difference ($t_1-t_2$) were histogrammed for calculation of average value spectrum and σ.

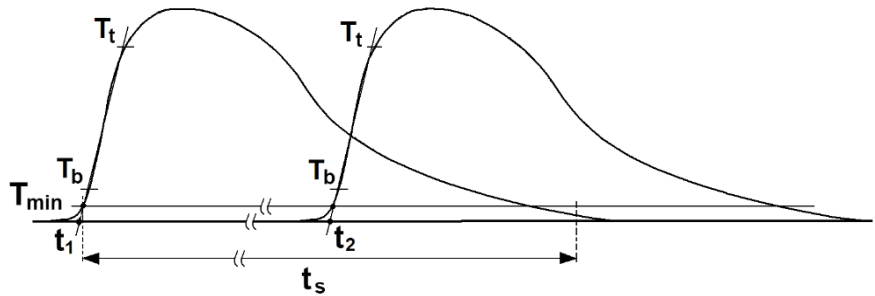

Fig. 2. Schematic representation of the algorithm used.

Comparison of signal parameters at registration of gamma quanta and electrons at different termination of TDT revealed the necessity for its complete termination, but in that case signal edges were changing up to ~ 15 ns at levels 0.1 - 0.9. The signals registered at different straw ends passed through the anode differently, which caused different changes of their leading edge incline. When moving the source from the straw center for 10-15 cm the changes were rather small, and with the growing distance the changes increased. Possible pulse shape changes caused by pulse superimposition with reflected signals at imperfect termination and by differences in the edge incline changes made it difficult to optimize the values of levels $T_b$ and $T_t$ for the entire length of the straw.

When registering electrons from $^{106}$Ru source, the straw termination was 360 Ω, gas gain was ~$8\times10^4$ and all the pulses were processed with no selection. The measurements along the straw were carried out with fixed levels $T_b$ and $T_t$ with values corresponding to 0.1 and 0.75 of the first pulse height, respectively. Curve 1 in Fig. 3 shows the longitudinal resolution σ varies from ~150 ps at the center to ~280 ps at the end of the TDT.

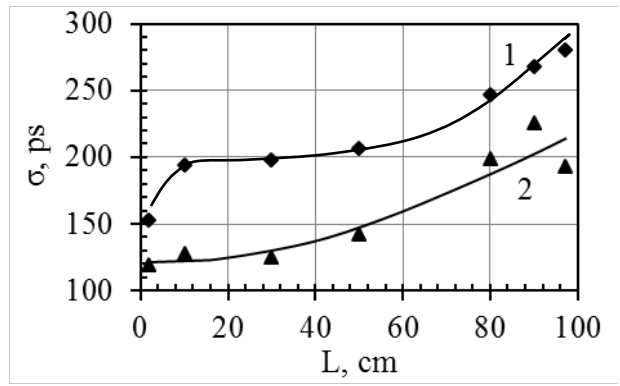

Fig. 3. The longitudinal resolution σ of straw depending on the position of the $^{106}$Ru source from the center of the straw. Curve 1 - resolution obtained for the fixed data. Curve 2 – resolution after adjustment of $\Delta N_i$ for each source position. Pressure gas mixture is 1 bar, gas gain is $8 \times 10^4$.

In order to improve the longitudinal resolution the dependence of $\Delta N_i$ along the anode was measured in units of the bins of the time axis, allowing for adjustment the measured values $\delta t_i$ by an algebraic summation of these values.

The dependence of the $\delta t_i$ adjusting shift in $\Delta N_i$ bins was determined by moving the source along the anode with 5 cm step (Fig.4, left). The best resolution was found for each point by shifting the measured $\delta t_i$ in its certain range. Using adjusting $\Delta N_i$ improves the longitudinal resolution. Curve 2 in Fig.3, found with use of adjusting $\Delta N_i$, shows the resolution along the straw in the range ~120 - ~230 ps. In comparison with Curve 1 the resolution has been improved by ~28%.

The correlation between the calculated time ($\delta t_c$) of pulse propagation of the distance from the anode center to the avalanche and the average measured value ($\delta t_m$) obtained with consideration of adjusting shift $\Delta N$ value was checked, which showed that the correlation existed and had to be taken into account. The correlation between the real and the measured times is presented in Fig.4 (right). It is seen that the measured $\Delta t_m$ values correlate to the real $\delta t_c$ time by linear correlation coefficient k=1.26. Consequently, the measured time value k×68.9, equals to 86.9 ps in our case, corresponds to 1 cm shift along the anode. Thus, the longitudinal resolution presented with curve 2 in Fig. 3 varies from 1.4 cm to less than 2.1 cm for 80% of the straw length and up to 2.6 cm for the full straw length.

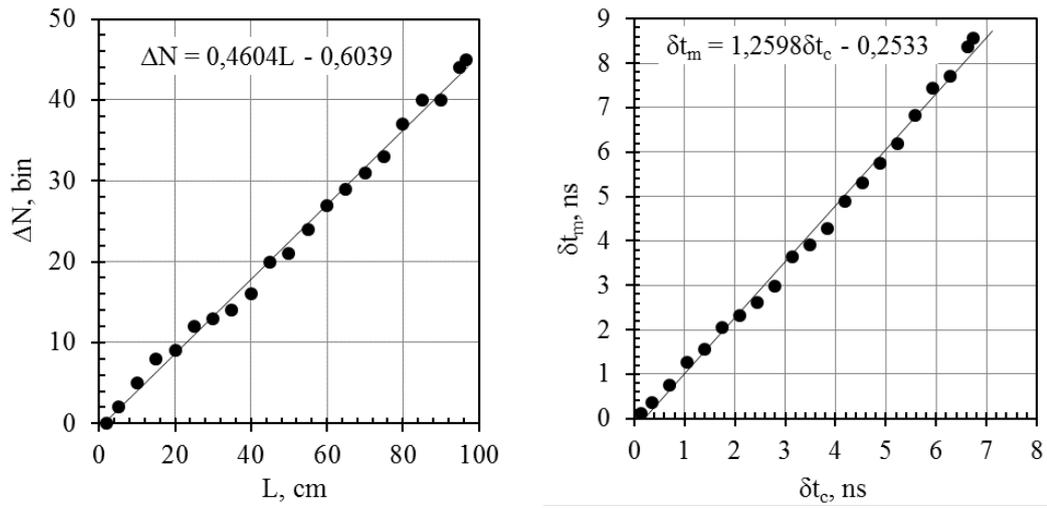

Fig. 4. Left: The value of the shift correction ΔN in the units of 200 ps bins depending on the distance from the center of the anode. Levels $T_b$ and $T_t$ are 0.1 and 0.75 of the first pulse height, respectively. The gas gain - $8 \times 10^4$.
Right: The correlation between the real $\delta t_c$ and the measured time $\delta t_m$.

When digitizing pulses that are analogues to registered with 2 m long straw at frequency of 5 and 2 GHz the value of time range $t_s$ (Fig.1) contains ~300 bins 200 ps in size and ~110 bins 500 ps in size, respectively. Obviously, when frequency decreases from 5 to 2 GHz the transmitted data volume is reduced by 2.7 times. The measurements carried out under the same conditions with 5 cm step of moving the $^{106}$Ru source showed the average value of the longitudinal resolution along the straw length deteriorated by ~20% with frequency reduction to 2 GHz (Fig.5). The worst resolution values were ~2.9 and 2.6 cm for 2 and 5 GHz, respectively.

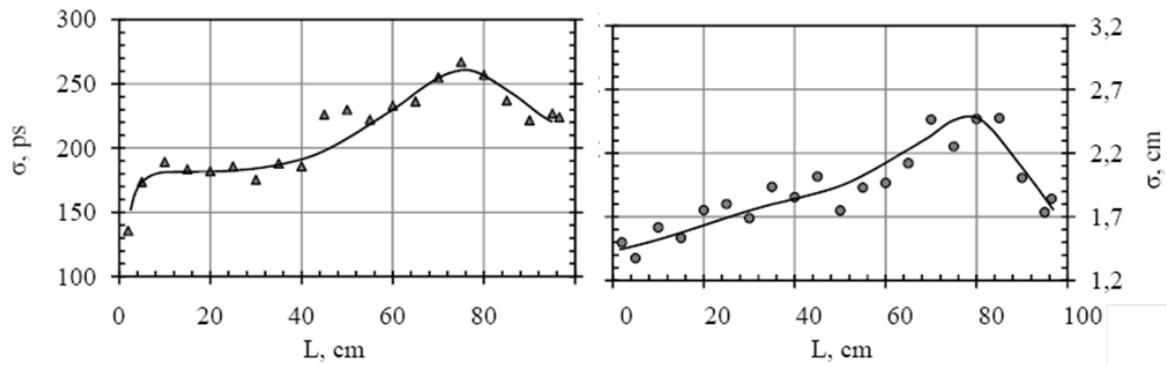

Fig. 5. Longitudinal resolution for the registration of the Ru-106 electrons at a frequency of 2 (left panel) and 5 GHz (right panel). Gas gain is $8\times10^4$. The levels of $T_b$ and $T_t$ are 0.1 and 0.7, respectively. Coincidence of the correlation coefficients is with accuracy of~ 1% .

The measuring technique with cable delay can be applied for estimation of the direct timing method possibility frontier in determining the longitudinal coordinate of straws. The longitudinal resolutions of TDT were measured during registration of pulses from $^{55}$Fe and $^{106}$Ru sources at a digitizing frequency of 2 and 5 GHz, the coefficients of linear correlations of about 1.24 and 1.25, respectively, were verified as well. The curves in Fig.6 show the values of the longitudinal resolution in the range from 100 to 130 ps along the straw, which corresponds to a resolution of 1.3 to 1.5 cm. The curves were obtained at a gas gain of $8 \cdot 10^4$.

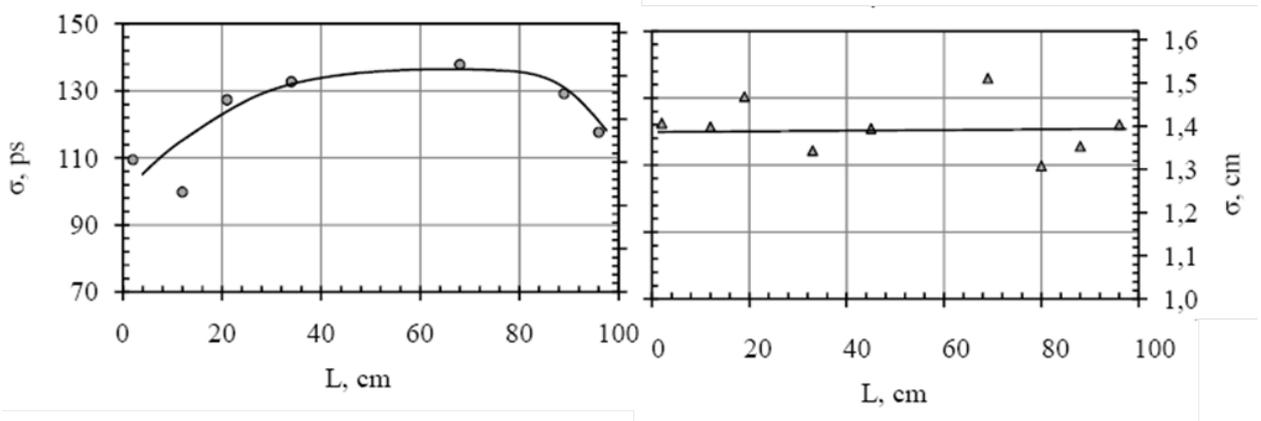

Fig. 6. The longitudinal resolution at a frequency of 5 GHz using a cable delay. Left panel: The source of Fe-55, the gas gain - $6 \times 10^4$, the levels of $T_b$ and $T_t$: - 0.1 and 0.7. Right panel: The source of Ru-106, the gas gain is $8 \times 10^4$, and the levels are 0.1 and 0.7.

### 4. Feasibility of using Constant Fraction Timing

The possibility of determining longitudinal coordinate by Constant Fraction Timing with use of DRS4 was verified for the read out at the two ends of straws. The method is based on the generation of bipolar pulse, that when crossing zero allows for definition $t_m$ of a time mark. The optimization for pulse timing is determined by the values of coefficients of inverted pulse delay ($t_d$) and direct pulse amplitude attenuation (f). The time difference between the two pulses $\delta t_m$ determines the position of the avalanche on the anode with respect to the center of straws. Effective fraction of amplitude $f_{eff}$ to which the time is referred in the case of the linear edges is defined by correlation $f_{eff} = f \cdot t_d / t_r (1 - f)$, where $t_r$ is the value of pulse edge.

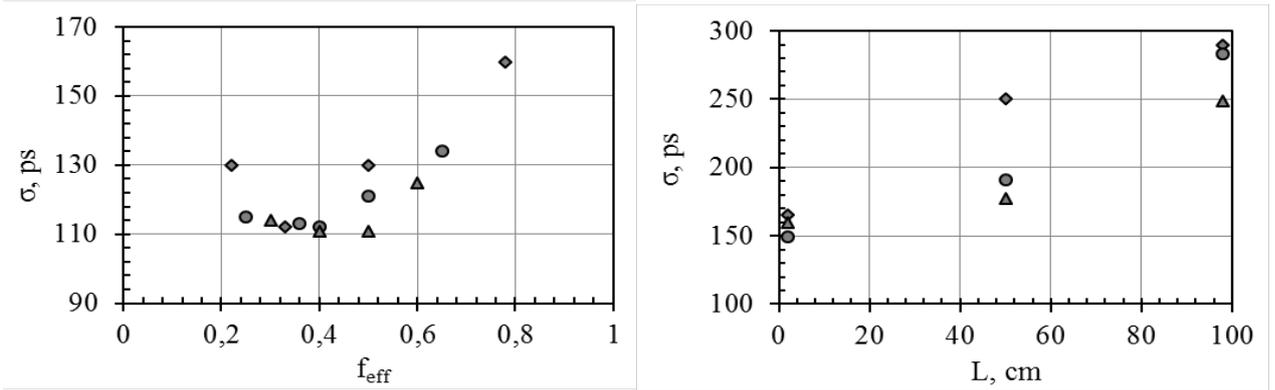

Fig. 7. Left panel: The time resolution in the center of the straw as a function of the level of discrimination $f_{eff}$ signals for different delay times $t_d$. Right panel: Longitudinal resolution along the straw in ps for different levels of binding $f_{eff}$.

The optimization of the method parameters ($t_d$ and f) was made under conditions of irradiation of the straw in its center by $^{55}$Fe source at different values of delay $t_d$. Fig.7 shows the resolution for several points along the straw. The calibration showed that the measured time difference between pulses equal to 90 ps corresponds to the 1 cm distance along the straw length. Therefore, spatial resolution for the method is ~1.7 cm in the center of the straw and decreases to ~ 2.9 cm at its end.

## 5. Summary

The possibility to measure longitudinal resolution in straws is of particular interest, especially being combined with maintaining readability of data on radial coordinate. The Charge-Division Technique possesses not only low longitudinal resolution, which quickly decreases with the straw length growth, but also lesser speed of operation in comparison with standard detectors with radial coordinate read-out. The Constant Fraction Timing is rather labour-consuming; it is not extensively used in multi-channel detectors. Determination of the longitudinal coordinate by the Direct Timing Method gives higher longitudinal resolution compared to the first two methods. The method can be applied to longer straws; it gives the resolution several times higher than it can be obtained by the Charge-Division Technique. It should be mentioned that the resolution of DTM is almost the same at registration of MIP and gamma-ray quanta from $^{55}$Fe source. The obtained longitudinal resolution values at registration of MIP for the 2 m long straw are better than 2.6 cm along its entire length (Fig.5). Estimation of the method possibilities shows that it can be improved up to 1.4 cm.

In order to obtain coordinate information by DTM along the straw, common parameters for the entire length of the straw are set; rapid current amplifiers are used for read-out, which are suitable both for the measuring time of electron drifts in the direction orthogonal to the anode and for measuring the difference of signal propagation times.

All of the aforesaid indicates feasibility of development of straw-based two-coordinate detectors with use of DTM, which may have sufficient speed of operation and do not seem to be too complicated.